\documentclass[aps,prl,twocolumn,floatfix,showpacs]{revtex4}
\usepackage{amsmath}
\usepackage{amsfonts}
\usepackage{amssymb}
\usepackage{graphicx}
\usepackage{color}

\begin{document}

\title{Phonon mediated electromagnetically induced absorption in hybrid opto-electro mechanical systems}
\author{Kenan Qu and G. S. Agarwal}
\affiliation{Department of Physics, Oklahoma State University, Stillwater, OK - 74078, USA}
\date{\today}

\begin{abstract}
We predict the existence of the electromagnetically induced absorption (EIA) in the double cavity configurations of the hybrid opto-electro mechanical systems (OEMS). We discuss the origin of the EIA in OEMS which exhibit the existence of an absorption peak within the transparency window. We provide analytical results for the width and the height of the EIA peak. The combination of the EIT and EIA is especially useful for photon switching applications. The EIA that we discuss is different from the one originally discovered by Lezama et al in atomic systems and can be understood in terms of the dynamics of three coupled oscillators (rather than two) under different conditions on the relaxation parameters. The EIA we report can also be realized in metamaterials and plasmonic structures.
\end{abstract}
\maketitle

The optomechanical systems have been recognized as good systems for the purpose of optical memories as the mechanical systems can have very long coherence times~\cite{network,router,coherence}. The realization that such systems can serve as memory elements became feasible by the prediction~\cite{EIT_GSA} of electromagnetically induced transparency (EIT) and the experimental demonstration of EIT by several groups~\cite{OMIT}. Much of this work was motivated by the corresponding work in atomic media~\cite{EIT}. While the EIT has been studied extensively in opto-electro mechanical systems (OEMS); a counter part of EIT namely the electromagnetically induced absorption (EIA) has not yet been investigated in OEMS. It may be noted that EIT is the result of destructive interference between different pathways leading to suppression of absorption. Thus one would think that there should be the possibility of constructive interference between different pathways. Such a possibility was first realized by Lezama and coworkers~\cite{EIA} in the context of atomic vapors. More recently EIT and EIA were demonstrated in plasmonic structures~\cite{plasmonic} where the interactions and phases  can be tailored by design of the structure thus enabling one to see either the EIT or the EIA behavior. As we discuss in this paper certain situations do warrant absorption or dispersion as was recognized by Harris and Yamamoto~\cite{abs1} and by Schmidt and Imam\u{o}glu~\cite{abs2}. In this paper we show the existence of EIA in a double cavity OEMS thereby filling a gap that has existed in the physics of OEMS. Such double cavity configurations are beginning to be studied in a number of papers for different applications~\cite{double}. We show how we can switch quite conveniently from EIT to EIA and vice versa by changing the power of the electromagnetic fields. The EIA that we discuss is different as we do not convert the transparency window into an absorption peak but we create an absorption peak in the transparency window. We note that several recent papers discuss a variety of new effects in double cavity OEMS. For example state transfer as well as squeezing using double cavity OEMS has been studied~\cite{double}. Further cavities with many mechanical systems are enabling one to reach very near quantum limit~\cite{qlimit}.

\begin{figure}[phtb]
 \includegraphics[width=0.40\textwidth]{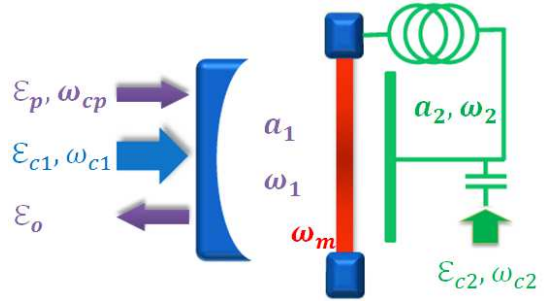}
 \caption{\label{Fig1}{(color online). Schematic illustration of the double cavity OEMS~\cite{double}. }}
\end{figure}

Lezama and coworkers~\cite{EIA} found that a simple three level lambda scheme cannot give rise to EIA. They considered optical transitions between the hyperfine states of atoms $F\to F'>F$ which showed the possibility of EIA. The work of Harris and Yamamoto~\cite{abs1} was based on a four level atomic shceme where one of the ground levels of the lambda scheme was connected by an optical transition to a higher level. This allowed the possibility of two photon absorption while at the same time suppressing one photon transition. Clearly if EIA were possible in OEMS, then we need to consider a more complicated configuration than, say, considered in the context of EIT: one needs to add an additional pump and at least one additional transition. Hence we study a double cavity configuration which is flexible enough to open up new pathways for the interaction with the probe field. We would show that the system of Fig.~\ref{Fig1} can produce EIA. We show how the EIA in our paper can be used to switch transition between two photonic routes in a manner similar to the Zeno effect used in several other types of systems~\cite{abs3}. The switching factor is very large (of the order of $3000$ in Fig.~\ref{Fig5}\textbf{a}).
We further show how EIA is useful in the transduction~\cite{Faust} of fields from optical to microwave domain. Our EIA is quite versatile as it would occur in metamaterials~\cite{meta,note2}, plasmonics~\cite{plasmonic,plasma} or in systems with several mechanical elements~\cite{qlimit}.

The double cavity configuration, as shown in Fig.~\ref{Fig1}, has become very popular recently~\cite{double}, and is becoming key to bring out very new features of the OEMS. The Hamiltonian for this system is given by
\begin{align}\label{1}
  H &= H_1 + H_2 + H_m + H_\text{diss}, \nonumber \\
  H_1 &= \hbar(\omega_1-\omega_{c1})a_1^\dag a_1 - \hbar g_1 a_1^\dag a_1 Q + \mathrm{i}\hbar\mathcal{E}_{c1}(a_1^\dag-a_1) \nonumber \\
   &\qquad + \mathrm{i}\hbar(\mathcal{E}_p a_1^\dag \mathrm{e}^{-\mathrm{i}\delta t} - \mathcal{E}_p^* a_1 \mathrm{e}^{\mathrm{i}\delta t}) \nonumber \\
  H_2 &= \hbar(\omega_2-\omega_{c2})a_2^\dag a_2 + \hbar g_2 a_2^\dag a_2 Q + \mathrm{i}\hbar\mathcal{E}_{c2}(a_2^\dag-a_2) \nonumber \\
  H_m &= \frac12\hbar\omega_m(P^2+Q^2),
\end{align}
and $H_\text{diss}$ is corresponding to the dissipation to the Brownian motion of the mechanical resonator with position $Q$ and momentum $P$ normalized, such that $[Q,P]=\mathrm{i}$, and the leakage of the photons from the cavity. Here $\omega_i$ is the resonant frequency of cavity $i$, and $\delta=\omega_p-\omega_{c1}$ is the detuning between the probe laser and the coupling laser in cavity $1$. The cavity $2$ is taken to be a microwave cavity. The coupling rate $g_i$ is defined by $g_i=(\omega_i/L_i)x_\text{zpf}$ with $L_i$ being the length of the cavity and $x_\text{zpf}=\sqrt{\hbar/(2m\omega_m)}$ being the zero point fluctuation for the mechanical resonator. The coupling fields and the probe field inside the cavities are given by $\mathcal{E}_{ci}=\sqrt{2\kappa_iP_i/{\hbar\omega_{ci}}}$ and $\mathcal{E}_p=\sqrt{2\kappa_1P_p/{\hbar\omega_1}}$, respectively. We employ a procedure which is now fairly standard in cavity optomechanics. We obtain the quantum Langevin equations and write the equations for the mean values. The mean value equations are solved around steady state by writing all expectation values in the form $A=\sum_{n=-\infty}^{+\infty}\mathrm{e}^{-\mathrm{i}n\delta t}A_n$. We obtain $A_n$'s perturbatively.  The cavity field $a_{i0}$ is the field in the $i$th cavity at the frequency of the coupling laser with frequency $\omega_{ci}$. The field $a_{i\pm}$ is the field at the frequency $\omega_{ci}\pm\delta = \omega_{ci}\pm(\omega_p - \omega_{ci})$, and more specifically $\omega_{1+}=\omega_p$. The output fields resulting from the applied probe field are defined as
\begin{align}\label{2}
    \mathcal{E}_{o1} &= 2\kappa_1(a_{1+}\mathrm{e}^{-\mathrm{i}(\omega_{c1}+\delta) t} + a_{1-}\mathrm{e}^{-\mathrm{i}(\omega_{c1}-\delta) t}) - \mathcal{E}_p\mathrm{e}^{-\mathrm{i}\omega_p t}, \nonumber \\
    \mathcal{E}_{o2} &= 2\kappa_2(a_{2+}\mathrm{e}^{-\mathrm{i}(\omega_{c2}+\delta) t} + a_{2-}\mathrm{e}^{-\mathrm{i}(\omega_{c2}-\delta) t}).
\end{align}
Note that the component $a_{2+}$ would yield the output at the frequency $\omega_{c2}+\omega_p-\omega_{c1}$ whereas the component $a_{2-}$ produces an output at the frequency $\omega_{c2}-\omega_p+\omega_{c1}$.
We would typically consider the situation when $\omega_p$ is close to the cavity frequency and the coupling field $\omega_{c1}$ is red detuned by an amount $\omega_m$. The fields $\omega_{c1}$ and $\omega_p$ combine to produce phonons at the frequency $\omega_p-\omega_{c1} \approx \omega_m$, with $Q_+\neq0$. This is the reason for the production of coherent phonons. We now concentrate on the output fields from the two cavities. We will display the numerical results for the normalized quantities defined by
\begin{equation}\label{3}
    \mathcal{E}_L = 2\kappa_1a_{1+}/\mathcal{E}_p, \qquad \mathcal{E}_R = 2\kappa_2a_{2+}/\mathcal{E}_p,
\end{equation}
which are fields at the frequency of the probe.
\begin{figure}[btp]
 \includegraphics[width=0.45\textwidth]{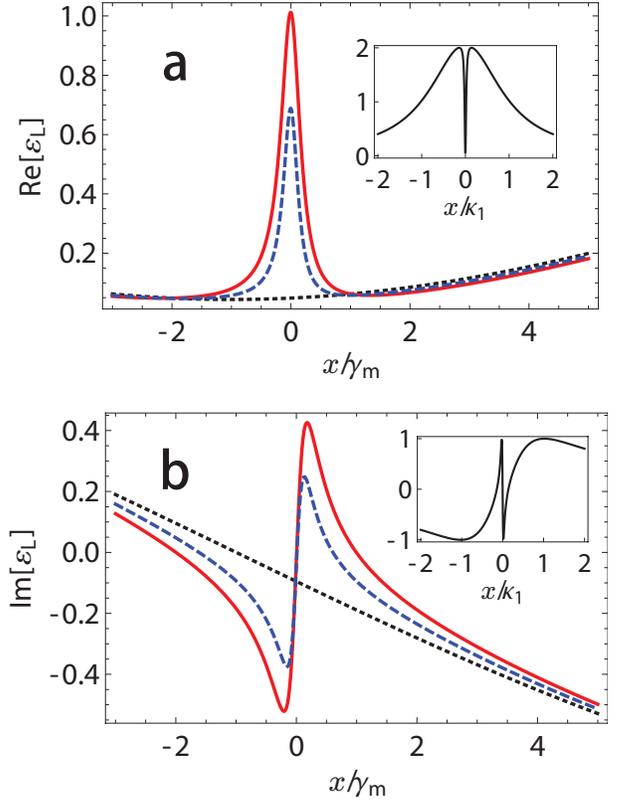}
 \caption{\label{Fig2}{(color online). The real (a) and imaginary (b) part of the field amplitude $\mathcal{E}_L$. The black dotted, blue dashed and red solid curves are corresponding to the cooperativity ratios $C_2/C_1=0,0.5,1$, respectively, and $C_1=40$. The response shows the effect of EIT when only one coupling field is present, and the emergence of EIA at the line center when both coupling fields are present. The insets show the EIT in a large frequency span with $C_2=0$, i.e. with no coupling field applied to the second cavity.}}
\end{figure}
The actual normalized output field from the cavity $1$ is given by $(\mathcal{E}_L-1)$, c.f. Eq.~(\ref{2}). We can find that the output is resonantly enhanced when $\delta\sim\Delta_1=\Delta_2=\omega_m$, where $\Delta_1=\omega_1-\omega_{c1}-g_1Q_0$ and $\Delta_2=\omega_2-\omega_{c2}+g_2Q_0$. In this regime, both the coupling fields are tuned by an amount $\omega_m$ below their corresponding cavity frequency, and the probe laser is in the vicinity of the cavity frequency $\omega_1$. We work in the resolved-side band regime $\omega_m\gg\kappa_{1,2}$. The detailed calculations lead to the following result for the output field $\mathcal{E}_L$
\begin{equation}\label{4}
    \mathcal{E}_L(x) = \cfrac{2\mathrm{i}\kappa_1}{(x+\mathrm{i}\kappa_1) - \cfrac{g_1^2|a_{10}|^2/2}{(x+\mathrm{i}\gamma_m/2) - \cfrac{g_2^2|a_{20}|^2/2}{(x+\mathrm{i}\kappa_2)}}},
\end{equation}
where $x(=\delta-\omega_m)$ denotes the detuning of the probe frequency to the cavity frequency. In what follows we assume that the probe field consists of many photons so that the effect of the thermal photons in the microwave cavity is negligible.

\begin{figure}[bpt]
 \includegraphics[width=0.48\textwidth]{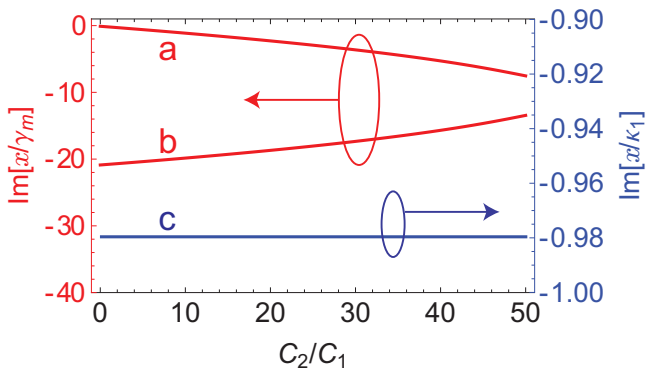}
 \caption{\label{Fig3}{(color online). The imaginary part of the roots of Eq.~(\ref{4}) with respect to the ratio $C_2/C_1$ of the cooperativities of the two cavities and $C_1=40$. Their real parts are almost zero. }}
\end{figure}
The structure of the output field $\mathcal{E}_L$ is very interesting. It shows how the resonant character of the output field changes from that of an empty cavity ($|a_{10}|=|a_{20}|=0$) to that of a single cavity ($|a_{20}|=0$) and to that of double cavities ($|a_{10}|\neq0,|a_{20}|\neq0$). The denominator in (\ref{4}) is linear in $x$ (an empty cavity), quadratic in $x$ (a single cavity), cubic in $x$ (double cavities). These changes determine the physical behavior of the OEMS. In order to see explicitly the nature of the output fields, we use the following set of experimentally realizable parameters $\omega_{c1}=2\pi\times4\times10^{14}$Hz,  $\omega_{c2}=2\pi\times10$GHz, $\omega_m=2\pi\times10$MHz, $\gamma_m=2\pi\times1$kHz, $\kappa_1=2\pi\times1$MHz, $\kappa_2=2\pi\times0.1$kHz, $g_1=2\pi\times50$Hz, and $g_2=2\pi\times5$Hz. We show in Fig.~\ref{Fig2} the numerical results for the two quadratures of the output field. These results clearly show the emergence of the EIA within the transparency window. The choice of the parameters to be used is dictated by the structure of (\ref{4}). We first note that for $|a_{20}|=0$, we have the standard EIT behavior (black dotted curves and the insets). We use a coupling power $P_{c1}$ below the critical power defined by $P_{cr}=\frac{\hbar\omega_l}{4g_1^2\kappa_1} (\kappa_1^2+\omega_m^2)(\frac{\gamma_m}{2}-\kappa_1)^2$, so that for $P_{c2}=0$, the two roots for $x$ are purely imaginary. The usual normal-mode splitting~\cite{NMS} occurs when the two roots have nonzero real parts, i.e. $P_{c1}>P_{cr}$. For the parameters above, $P_{cr}\approx 16.6$mW. For $P_{c1}<P_{cr}$, the interference then leads to the EIT window with a width $\Gamma_\text{EIT}=(1+C_1)\gamma_m/2$, where $C_i=g_i^2|a_{i0}|^2/\kappa_i\gamma_m$ denotes the optomechanical cooperativity of cavity $i$. For the chosen parameters and for $P_{c1}\approx1.3$mW, $C_1=40$. Clearly, if we want to produce an absorption peak within the EIT window, then we need to choose $C_2$ such that the third root of the denominator in Eq.~(\ref{4}) lies within the EIT window. For the results shown in Fig.~\ref{Fig2}, we choose $C_2=C_1/2$ (blue dashed curves) and $C_2=C_1$ ( red solid curves), corresponding to which $P_{c2}\approx1.6\mu$W and $3.3\mu$W. We show in Fig.~\ref{Fig3} how the roots of the denominator in Eq.~(\ref{4}) change for $P_{c1}$ below the critical power and if the driving field in the cavity $2$ is increased. For $C_2=0$, the width of the EIT window is $20.5\gamma_m$. The curve \textbf{c} gives the overall width within which the transparency window appears. The curve \textbf{a} gives the width of the EIA peak within the EIT window.
We now examine quantitatively the width of the absorption peak. When $|a_{20}|^2=0$, $P_{c1}<P_{cr}$, the two roots of the denominators in (\ref{4}) are $\kappa_1$ and $\Gamma_\text{EIT}$, and $\Gamma_\text{EIT}\ll \kappa_1$. In presence of the additional coupling field $a_{20}\neq0$, the root $\Gamma_\text{EIT}$ splits into two parts
\begin{align}\label{5}
    & \Gamma_\text{EIT} \to \Gamma_\pm = \frac12\Gamma_\text{EIT} \pm \frac12\sqrt{\Gamma_\text{EIT}^2-2g_2^2|a_{20}|^2}, \nonumber\\
    & \Gamma_- = \Gamma_\text{EIA} \approx \kappa_2 + \frac{g_2^2|a_{20}|^2}{2\Gamma_\text{EIT}}, \quad \text{if } \frac{2g_2^2|a_{20}|^2}{\Gamma_\text{EIT}^2}\ll 1 .
\end{align}
The existence of an additional splitting in the roots $\Gamma_\pm$, especially when $\kappa_2\ll\Gamma_\text{EIT}$, leads to the absorption peak within the transparency window. The half width of the absorption peak is given by $\kappa_2 + g_2^2|a_{20}|^2/2\Gamma_\text{EIT}$. It should be borne in mind that the microwave cavity is especially useful as $\kappa_2\ll\gamma_m,\Gamma_\text{EIT}$. The root $\Gamma_-$ has the behavior given by the curve \textbf{a} in Fig.~\ref{Fig3}.

\begin{figure}[tpb]
 \includegraphics[width=0.45\textwidth]{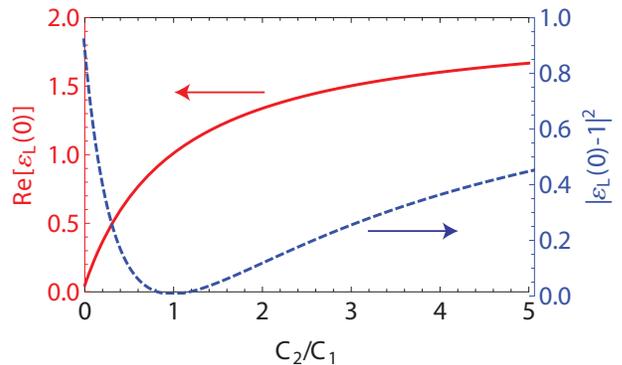}
 \caption{\label{Fig4}{(color online). This response of the double cavity OEMS under the effect of coupling fields in both cavities. The solid red curve illustrates the real part of the field amplitude inside the optical cavity at its line center, while its imaginary part is $0$. The dashed blue curve illustrates the intensity of the output field from cavity $1$ at its line center. }}
\end{figure}
We will now study the characteristics and the origin of the EIA peak. From Eq.~(\ref{4}), we get the height of the EIA peak, $\mathcal{E}_L(0)\approx 2/(1+C_1/C_2)$. Note that the height of the EIA peak depends on the ratio of the cooperativity parameters $C_i$ for the two cavities. We exhibit the behavior of the EIA peak and the output field at the probe frequency $|\mathcal{E}_{o1}(\omega_p)|^2/|\mathcal{E}_p|^2 = |\mathcal{E}_L-1|^2$ as a function of the ratio of the cooperativity parameters in Fig.~\ref{Fig4}. Notice from this figure that we get perfect EIA when the ratio of the two cooperativity parameters is unity. At this point, the probe field emerges from the second cavity as displayed in Fig.~\ref{Fig5}. The Fig.~\ref{Fig5}\textbf{a} clearly shows how the route of the probe photons changes by the increased absorption resulting from the coupling to the second cavity. This is analogous to the idea of using the Zeno effect~\cite{abs3}, i.e. increasing decoherence to switch the path of the photon. According to the procedure outlined after Eq.~(\ref{1}), the probe field produces steady state of the mechanical mode as $(Q_+\mathrm{e}^{-\mathrm{i}\omega_mt} + Q_-\mathrm{e}^{\mathrm{i}\omega_mt})$. The Fig.~\ref{Fig5}\textbf{b} shows the behavior of the mechanical mode which goes from a bright mode to an almost dark mode when $C_1=C_2$. We have concentrated on our demonstration of EIA within the transparency window, though it is possible to have EIA for other ranges of parameters.
\begin{figure}[tpb]
 \includegraphics[width=0.45\textwidth]{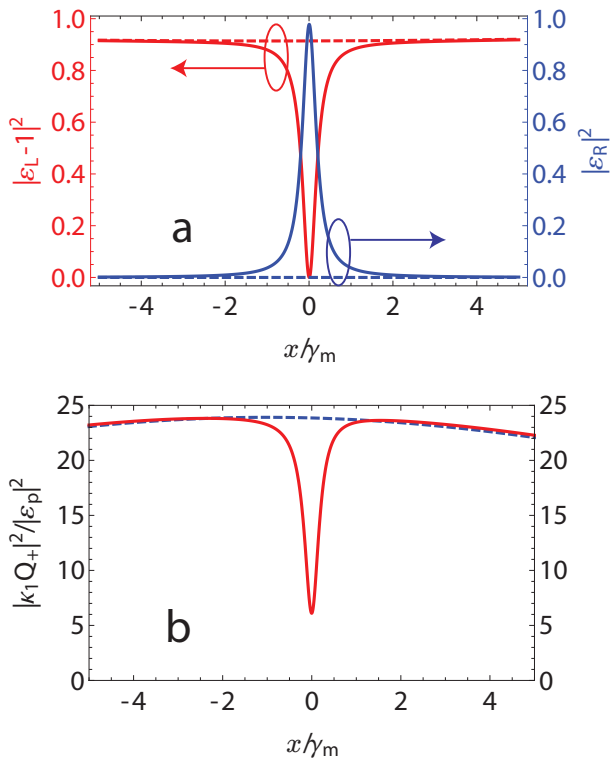}
 \caption{\label{Fig5}{(color online). (a) The normalized output from the first cavity $|\mathcal{E}_L-1|^2$ and from the second cavity $|\mathcal{E}_R|^2$; and (b) the amplitude of the mechanical displacement normalized to $|\mathcal{E}_p|^2$. The system behaves (almost) perfect reflection with bright mechanical mode when $C_2=0$ (dashed curves); and it behaves perfect transmission with nearly dark mechanical mode when $C_2=C_1$ (solid curves). }}
\end{figure}

We next present a coupled oscillator model which shows the existence of EIA. Note that the coupled oscillator models can very often mimic a variety of physical phenomena. In fact two coupled oscillators~\cite{analog} have been used to understand EIT as well as EIA~\cite{plasmonic,EIA}. It turns out that the EIA of the type discussed in this paper has to be understood in terms of three coupled oscillators --- in our case two of these ($u$ and $v$) would represent cavity modes and the third one ($w$) the mechanical oscillator. The three effective oscillators can be described by equations (written in rotating wave approximation) as
\begin{equation}\label{6}
    \begin{aligned}
        \dot{u} &= -\mathrm{i}\Delta_1 u - \mathrm{i}G_1w - \kappa_1u + \mathcal{E}_p\mathrm{e}^{-\mathrm{i}\delta t}, \\
        \dot{v} &= -\mathrm{i}\Delta_2 v - \mathrm{i}G_2w - \kappa_2v, \\
        \dot{w} &= -\mathrm{i}\omega_m w - \mathrm{i}G_1u - \mathrm{i}G_2v - (\gamma_m/2)w. \\
    \end{aligned}
\end{equation}
These three coupled equations can exhibit a variety of phenomena depending on the couplings $G_1$, $G_2$ and the relaxation parameters $\kappa_1$, $\kappa_2$ and $\gamma_m$. For the existence of the EIA, it is simple to have $\kappa_1\gg\gamma_m\gg\kappa_2$. Note that a whole class of hybrid systems coupling optical and microwave systems can be described by Eqs.~(\ref{6}) and their quantum version in terms of Langevin equations~\cite{zhu}.

In conclusion, we have demonstrated the possibility of the EIA within the transparency window of the optomechanical systems. For the OEMS of this paper, the EIA results in the transduction of optical fields to microwave fields. Note however that the transduction of fields at single photon levels would require a full quantum treatment as in~\cite{router}, although quantum ground state is now realized~\cite{Teufel}. The EIA within the transparency window is quite generic and is applicable to a variety of systems, which can be effectively described by three coupled oscillators. These systems would include other types of optomechanical systems like those containing two mechanical elements~\cite{qlimit,sumei}, two qubits~\cite{qubits}, or very different classes of systems like plasmonic structures~\cite{plasmonic,plasma} and metamaterials~\cite{meta,note2}.

G.S.A. would like to thank the Director of Tata Institute of Fundamental Research, Mumbai, where a part of this work was done.

\end{document}